\documentstyle[preprint,eqsecnum,aps]{revtex}

\begin{document}
\draft
\title{Enhancement of localization length for two interacting kicked rotators}
\author{Fausto Borgonovi\\}
\address{
Dipartimento di Matematica, Universit\`a Cattolica,
via Trieste 17, 25121 Brescia, Italy  \\
Istituto Nazionale di Fisica Nucleare, Sezione di Pavia,
via Bassi 6, 27100 Pavia, Italy.
}
\author {Dima L. Shepelyansky$^{*}$}
\address {
Laboratoire de Physique Quantique, Universit\'{e} Paul Sabatier,\\
118, route de Narbonne, 31062 Toulouse, France
}

\date{9 March 1995}
\maketitle
\begin{abstract}
We study the effect of coherent propagation of two interacting particles
in a disordered potential. The dependence of the enhancement
factor for coherent localization length due to interaction
is investigated numerically in the model of quantum chaos.
The effect of interaction for two particles in many dimensions is also
discussed.
\end{abstract}

\pacs{
\hspace{2.9cm}
}


\newpage
\section{Introduction}
\label{sec:level1}

The quantum localization of dynamical chaos has received a great
deal of attention during the last years \cite{Boris}, \cite{Felix}.
It has been understood that quantum interference effects
lead to a suppression of diffusive spreading in the action
space in spite of the chaotic dynamics of the correspondent classical
model.
An important consequence of this phenomenon  is the exponential localization
of quantum eigenfunctions over the unperturbed levels.
A close correspondence has been established between this
dynamical localization and the Anderson localization in
a random potential for solid state  systems \cite{Fish}.
One of the most studied models in this field is the
kicked rotator model  (KRM), which in  the classical limit
corresponds to the Chirikov  standard map\cite{Boris1} (CSM), a common
paradigm of classical chaos.
Although KRM seems to be at a first glance a pure mathematical model,
it has however
found important applications for real physical systems, as for example
the process of microwave ionization of Rydberg atoms \cite{Dimav}.
Another useful property of KRM is that it can be studied very
efficiently in numerical simulations allowing to investigate its
properties in great details.

Anyway KRM describes the one particle quantum  dynamics and in many
respects it is quite similar to
one particle  localization   in a quasi one-dimensional random
potential \cite{RevMod}.
The last problem has been intensively studied in solid state physics
and it is well understood from the theoretical point  of view.
On the contrary the case of interacting particles is much more complicated
and a clear theoretical picture is still lacking (see for instance
the recent Review \cite{Kirk}).
Usually this problem is studied near the ground state
and it is a common lore   that a repulsive
interaction would result  in a even stronger   localization
\cite{Tieri,DFisher}.
However, the recent investigation of two interacting particles
in a quasi-one-dimensional
random potential \cite{TIP} has shown that, even
in the case of repulsive particles, interaction leads to an
enhancement of localization length and a coherent propagation of two particles
on a distance $l_c$ much larger than the one-particle
localization length $l_1$. According to \cite{TIP} the enhancement
factor is given by~:
\begin{equation}
{l_{c} \over {l_1}} \sim {l_1 M} { U^2\over {32 V^2}}
\label{est}
\end{equation}
where $U$ is the strength of on site interaction between two
particles, $V$ is the one-particle hopping element between
nearest sites which determines the size of one-particle energy
band and $M$ is the number of transverse channels so that
by itself $l_1 \sim M$ ($M < l_1$). In (\ref{est}) the intersite distance $a$
is
taken to be equal one and energy is taken near the center of the band
so that $k_{F} \sim 1/a=1$. Interaction $U$ is considered to be
less or comparable with $V$. The equation (\ref{est}) is valid in the
regime when the enhancement factor is larger than one $l_c/l_1 > 1$.

The physical reasons according to which two particles propagates together,
forming an effective pair,
can be understood in the following way \cite{TIP}. Since
the interaction couples only nearby sites (or on site)
then the particles initially located on a distance $r \gg l_1$
have only exponentially small effective coupling with each other
due to the exponential decay of localized one-particle eigenstates.
For such type of states two interacting
particles (TIP) remain localized near their initial positions.
These states are localized and form the majority of all states.
However, there are other states in which TIP
are initially close so that $r < l_1$. As it follows from the
previous case $r \gg l_1$, the initially close particles ($r < l_1$)
cannot become separated on a distance much larger than $l_1$
(otherwise we would contradict the previous case
$r \gg l_1$).
Therefore, they always remain on a distance $r \sim l_1$.
On this distance interaction between particles is important.
Qualitatively, the motion of one particle with respect to the other
produces some kind of noise on it. This
noise gives destruction of interference effects which
have produced one-particle localization. The destruction of
interference allows particles to propagate on a distance $l_c$
which is much larger than the distance between particles $l_1$.
This propagation can take place only for two particles since as soon as
they become separated ($r \gg l_1$) they become localized.
In some sense the way of TIP propagation in a random potential
is similar to the method used by M\"{u}nchhausen to save himself from a swamp.

The functional dependence in (\ref{est}) can be understood
 in the following way.
The enhancement factor $l_c/l_1$ is proportional
to the probability to mix one-particle states by interaction.
In fact $l_c/l_1 \sim \Gamma \rho$ where $\Gamma \sim {U_s}^2 \rho$ is the rate
of transition and $\rho$ is the density of coupled states \cite{TIP}.
The coupling matrix element is $U_s \sim U/{N}^{3/2}$ \cite{TIP}
where $N=M l_1$ is the number of unperturbed components
in one-particle eigenstate. The density of states is
$\rho = 1/{\Delta E} \sim {N^2}/V$ so that the enhancement factor is
$l_c/l_1 \sim (U_s/\Delta E)^2 \sim N (U/V)^2$ in agreement with (\ref{est}).
It is interesting
to mention that such type of estimate is quite similar to the derivation of
statistical enhancement for weak interaction and parity
violation in neutron-nucleus reactions discussed by
Sushkov and Flambaum \cite{Sushkov}. However
in \cite{Sushkov}, even enhanced, the interaction remains relatively weak
giving only small corrections while in our case the enhanced interaction
$U_{eff} \sim (M l_1)^{1/2} U$ is not small and it leads to significant
changes in the properties of the system.

Another way of derivation of (\ref{est}) is based on the reduction of the TIP
problem to a problem of superimposed band random matrices (SBRM) \cite{TIP} .
There, the interaction, even if repulsive, creates
an effective thick wire
along the diagonal $n_1=n_2$
in the two-dimensional plane $(n_1,n_2)$ of indices corresponding
to two particles. The effective width of the wire (effective
number of transverse channels) is
determined by the number of levels
$M l_1$ coupled by interaction in one-particle basis.
Outside of this width interaction is exponentially small
and  can be neglected at least in the first approximation.
The large number of effective transverse channels $M_{eff} = M l_1$ leads to
enhancement of localization length with the enhancement
factor proportional to $M_{eff}$.

All the above approaches were based on the assumption of
statistical independence of transition matrix elements and
eigenenergies in one-particle basis. This approximation
seems to be reasonable due to randomness of potential and
finite radius of interaction. However, it is very important to
have a direct check and to verify the prediction (\ref{est}).
Some numerical checks were presented in \cite{TIP}.
Here we present the results of a more detailed
numerical investigation which we carried out for the model of
interacting kicked rotators which had been discussed in \cite{TIP}.
We also present numerical results for a model with
finite radius of interparticle interaction.

     The paper is constructed as follows. In  section II we
introduce the model and present the main results for on site
interacting kicked rotators. The case of finite radius of interaction
is discussed in  section III. Conclusions and discussions
of results are presented in section IV.

\section{The ``on site'' interaction model}
\label{sec:level2}

To investigate the effect of enhancement of localization length
by interaction we used the model of two interacting kicked rotators
introduced in \cite{TIP}. The model represents two particles
on a ring perturbed by kicks periodic in time. The evolution
of the wave function $\psi$ on one
period of perturbation is described by the unitary operator (Floquet
operator)~:
\begin{equation}
\hat {S} = e^{-iT(\hat{n}_1^2+\hat{n}_2^2)/2+ iU\delta_{n_1,n_2}}
\times e^{-ik(\cos\theta_1+\cos\theta_2)}
\label{FLOQ}
\end{equation}
with ${\hat n}_{1,2}=-i {\partial}/{\partial {\theta_{1,2}}}$. For $U=0$
we have two noninteracting kicked rotators which had been
intensively studied during the last years \cite{Boris}, \cite{Felix},
\cite{Fish}.
The classical dynamics is chaotic and diffusive
for the chaos parameter $K = kT > 1$ \cite{Boris1}.
The diffusion rate is approximately $D=n^2/t = k^2/2$ for
$K \gg 1$. Quantum
interference effects lead to suppression of this diffusion
for typical irrational values of $T/4\pi$
and to exponential localization of eigenstates so that the
averaged probability
distribution over unperturbed levels decays as
${\mid \psi_{n} \mid}^2 \approx \exp(-2 \mid n -n_0 \mid/l_1)/l_1$.
The localization length in the region of strong chaos is approximately
given by $l_1 \approx D \approx k^2/2$. Quasiclassical regime corresponds to
$k \gg 1, T \ll 1, kT= const$ and $l_1 > 1$.

For
$U\ne 0$ interaction between particles is switched on.
Using the Bessel expansion, eq. (\ref{FLOQ}) can be written as~:
\begin{equation}
\hat {S} = e^{-iT(\hat{n}_1^2+\hat{n}_2^2)/2 +iU\delta{n_1,n_2}}
\times \sum_{m_1,m_2} J_{n_1-m_1} (k) J_{n_2-m_2}(k) (-i)^
{n_1+n_2-m_1-m_2} e^{im_1\theta_1+im_2\theta_2}
\end{equation}
Since $J_{ n - m} (k)$
is exponentially small when $\vert n - m \vert > k$,
at each iteration of $\hat S$ many states ($\sim 2k$) are coupled.
Inter--particles interaction acts only when the two particles
have the same momentum,
namely when they occupy
the same site on the momentum grid, on site interaction,
if we adopt the solid state terminology.
Due to the  presence in the exponent, the interaction $U$ can only
take values in the interval $(0,2\pi)$.
Due to interaction the two particles are able to propagate
coherently on a distance $l_c$ much larger than the original
one particle localization length $l_1$, as was anticipated
in the introduction. Of course this can happen if they are
initially started within a distance $r  < l_1$.
Even if very close with the TIP problem in 1d Anderson model
\cite{TIP},
our model has three different features. Indeed no randomness is here acting
and the interaction is neither attractive, nor repulsive.
In addition the perturbation   couples many levels at each iteration
(kick).

The quantum dynamics was investigated in numerical
simulations for symmetric configurations
with an effective number of unperturbed levels from $1000$ to
$2000$. Antisymmetric configurations of two particles do not
feel the on site interaction $U$ and are not interesting.
We iterated the quantum operator $\hat {S}$ starting from
two particles initially at the same site for different
parameters values. The spreading of the wave function in the 2-d space
$(n_1,n_2)$ was studied  through the second moments along
the diagonal line $n_1=n_2$~:
\[
\sigma_{+} (t) = {1\over 4} \langle ( \vert n_1\vert +
\vert n_2 \vert )^2 \rangle_t
\]
and across it
\[
\sigma_{-} (t) = \langle ( \vert n_1 \vert - \vert n_2 \vert )^2 \rangle_t
\]
as a function of the iteration time $t$.
In any investigated case $\sigma_+$ was observed to saturate at an
higher value than in absence on interaction, see Fig.1.
On the other side $\sigma_{-}$ keeps the same order of
magnitude as $l_1^2$ (as it should be in absence of interaction).
This means that the localization length is strongly enhanced
along the diagonal $n_1=n_2$ while it remains localized, with
roughly  the
same localization length, across the diagonal.
This is even more evident if one looks on the probability
distribution $P(n_1,n_2) = \vert \psi(n_1,n_2)\vert^2$ at a
fixed time $t \gg t^{*}$, where $t^{*} \approx l_1$
is the localization time, see for instance Fig.2.
In this picture a local averaged distribution function is
represented in the quarter of space $n_1,n_2 >0$, in a semilog
plot. The   channel of propagation along the
diagonal $n_1=n_2$ is manifested
in the contour lines drawn at the surface basis.

 From the distribution function important information can be extracted
by computing the following distributions~:
\[
P_{\pm} (n_{\pm}) = \sum_{\vert n_1 \pm n_2 \vert = n }
\vert \psi(n_1, n_2)\vert^2
\]
represented in Fig.3 as a function of
$n_{\pm}= \vert n_1 \pm n_2 \vert/\sqrt{2} =n/\sqrt{2}$. These distributions
give a measure of the ``perturbed'' localization lengths along
$(+)$ and across $(-)$ the principal diagonal.
It is relatively easy to derive from them the respective localization
lengths $l^{\pm}$
by the usual best fitting procedure.
Indeed the distributions are quite close
to  exponential curves $P_{\pm} \sim 2 \exp(-{2n_{\pm}/l^{\pm}})/l^{\pm}$
as can be inferred from Fig.3.
In the same picture we show the probability distribution in absence
of interaction $P_{+}^0 = 8 n_{+} \exp(-2^{3/2}n_{+}/l_1)/{{l_1}^2}$
with $l_1=k^2/2$. This noninteracting distribution is
quite similar to $P_-$.
The localization lengths $l^{\pm}$ are then plotted vs $l_1=k^2/2$
in order to check the validity of Eq. (\ref{est}).
For sake of comparison the lines with power 1 and 2 are drawn. The
dependence of coherent localization length
$l_c=l^+$ on one-particle length $l_1$ can be satisfactory
described by $l^+ \approx 0.5 {l_1}^2$ at $U=2$ while
$l^- \approx 1.5 l_1$.
However, the least square fit for the data of Fig.4
gives $l^{\pm} \sim {l_1}^{\alpha_{\pm}}$ with
$\alpha_+ = 1.44 \pm 0.29$ and $\alpha_- = 1.14 \pm 0.07$.
We attribute the difference from the theoretical values
2 and 1 to the insufficiently large interval of variation
of $l_1$ (only 4 times).  Further more detailed numerical
investigations should be done to extract more accurate values of
$\alpha_{\pm}$. Another interesting point following from the
Figs.1-4 is that for the same length $l_1$ the
coherent length $l_c$ is significantly larger than in
TIP in 1d Anderson model considered in \cite{TIP}.
This can be seen by direct comparison of  $\sigma_{+}$ values.
One of the reasons for this difference could  be the different
type of hopping
in KRM where one kick couples many levels.

To determine the numerical factor in the dependence of $l_c$
on both $U$ and $l_1$ one should also study the problem
at small values of $U \ll 2$. However, here for observation
of the enhancement $l_c/l_1$ one should work at much
larger values of $l_1$ than we used in Figs.1-4. This requires
a sharp increase of the basis and makes the numerical calculations
too difficult. Therefore, to investigate the dependence on $U$ we
did the following.
According to (\ref{est}) we expect
that it should exist a critical $U_{cr}$ given by $U_{cr} \sqrt{l_1}>C$
with $C \sim 1$.
To check this we consider the same model but with random rotating phases,
which means that $T(n_1^2+n_2^2)/2$ in the first
exponent is replaced by $f(n_1)+f(n_2)$ with $f(n)=f(-n)$
being a random function  in the interval $(0,2\pi)$.

In this way we can change configuration varying the random realization
and obtaining results for the average behaviour.
The results averaged
over 10 realizations of disorder are presented in Fig.5.
The asymptotic value reached by
the second moments
$\sigma_{\pm}^\infty =\lim_{t\to \infty} \sigma_{\pm} (t)$
are plotted in units of the same value in absence of interaction ($U=0$).
Error bars are due to fluctuations in varying the random configuration.
For small $U$, $\sigma_+^\infty$ and $\sigma_-^\infty$ are both
increasing up to double their  value  without interaction.
When $U > U_{cr}$ full ( $\sigma_+$) and open ($\sigma_-$)
circles start
to deviate one from each other thus indicating the presence
of a sharp transition.
In our case the
transition starts at
one particle localization length $l_1 = 8$
which approximately agrees with the observed critical value
$U_{cr} \approx 0.3$ and $C \approx 1$.
We were not able to extract a more precise information on
the dependence of $l_c$ on $U$ due to the  heaviness of numerical
simulations.

\section{The model with finite radius of interaction}
\label{sec:level3}

In this section we analyzed the effect of a finite range interaction
on the dynamics. To be more precise we chose in (\ref{FLOQ}),
instead of the former
on site interaction $U\delta_{n_1,n_2}$ a more general,
finite radius interaction~:
\[
U \eta (n_1 , n_2) \theta (b- \vert n_1 - n_2 \vert)
\]
where $\theta (x)$ is the usual step function which is zero for $x<0$
and one for $x \geq 0$. The phase $\eta$ is a random number in
the interval $(-1,1)$
which depends only on $n_1$ if $n_1<n_2$ and only on $n_2$ if
$n_2 < n_1$.
It is quite clear that for $b=0$ it becomes the previous one with
diagonal disorder. The diagonal disorder creates some difference from the model
of section II since now the interaction depends not only from
the difference $n_1 - n_2$. However, physically it is clear
that diagonal disorder in interaction will not change too much the
results. Indeed, the main point is to have {\it some} coupling between
two particles and the sign of interaction is not very important for
the destruction of interference, since
the one-particle random potential is already acting.
Our numerical results confirm
that disordered one site interaction gives qualitatively the same
effects as for on site interaction $U\delta_{n_1,n_2}$.
We usually investigated the cases with different
interaction radius $R=2b+1$ and $U=\pi$.

Our main interest is to investigate the effect of
interaction with finite
radius $R $. From the theoretical point of view we
can expect that for interaction radius $R < l_1$ the equation
(\ref{est}) is still valid since the particles are
effectively coupled on a distance $l_1$. However,
for $R > l_1$ the size of the effective thick wire on the lattice
$n_1,n_2$ is defined by $R$ so we can expect that
the enhancement factor will become
larger $l_c/l_1 \sim (R+l_1)$. This expression should remains valid up
to values of $R \ll l_{12}$ where $l_{12}$ is the 2d localization
length for infinite radius $R$: $\ln l_{12} \sim l_1$.
Indeed, for $R \gg l_{12}$ with the chosen type of interaction
one should have the same localization length as in 2d.

The results of our numerical simulations for finite
interaction radius $R$ are presented in Figs.6-9.
In Fig.6 and Fig.7 we show $\sigma_+$ and $\sigma_-$ for three
different $b$ values $(b=0, 4, 16)$ which roughly agree
with the above estimates.
In agreement with the above picture the enhancement factor
remains practically unchanged for $R < l_1$.
Only the case $b=16$ has $R > l_1$
and this produces a significant growth in $\sigma_+$ (and even in
$\sigma_-$). The distributions $P_{\pm}(n)$, as defined
in the previous section, are shown in Fig.8
and demonstrate a sharp increase of $l^+$ comparing to $U=0$.
In the same way we took the asymptotic values
$\sigma^\infty_{\pm}$ reached by
$\sigma_{\pm}(t) $ at large time $t$
and we plot in Fig.9 their square root as a function of the radius of
interaction $R$ (the values of $l^{\pm}$ have
a similar behaviour).
This figure confirms the above arguments that the enhancement
starts to grow only for $R > l_1$.
However, it should be mentioned that the increase of $R$
leads to a growth not only of $\sigma_+$ but also of $\sigma_-$.
Indeed, $\sigma_+$ and $\sigma_-$ are growing in the same way:
from the same Fig.9 one can see that the ratio
$\sigma_{+}^{\infty}/\sigma_{-}^{\infty}$ is approximately
constant as a function of the interaction radius $R$.
The physical explanation of this similar growth is quite simple:
the increase of $R$ leads not only to the increase
of coherent propagation length but also to an increase of the
effective size of the pair which becomes of the order of $R \gg l_1$.
Unfortunately, we were not able to study numerically the regime
$R \gg l_1$ (in our case the maximal ratio $R/l_1 \approx 2$)
and it was not possible to check the dependence $l_c \sim R l_1$.

\section{Conclusions and discussions}
\label{sec:level4}

Above we presented the results of our numerical investigation
about two interacting kicked rotators in the domain of quantum chaos.
They clearly demonstrate that on site interaction
between two rotators in momentum space leads to large enhancement
for localization length comparing to noninteracting case (Figs. 1-3).
The localization length for coherent propagation of
two particles $l_c=l^+$ is significantly larger than the distance
between them $l^{-} \approx l_1$. The maximal ratio
$l^{+}/l^{-}$ in our numerical simulations was near 10
(Fig.4) which justifies the fact of effective
enhancement of localization length for coherent propagation of two
particles. The direct check of the relation (\ref{est}) shows that
the coherent localization length $l_c=l^+$ grows approximately
as $l^+ \sim {l_1}^2$ but  more detailed
numerical calculations are necessary to have a more accurate
check of the power (see also  discussion below).

Another part of our investigations was devoted to the effects of
a final radius of interaction $R$ between particles. They definitely
show that for $R < l_1$ the enhancement is not sensitive to the
value of $R$ (Fig.9).  The physical reason is quite clear.
Indeed on site interaction couples one-particle states in a radius
of $l_1$  and therefore interaction with $R < l_1$ does not give
significant changes. For $R \gg l_1$  the enhancement factor
starts to grow with $R$. One can expect that in the regime
$R \gg l_1$ the radius $R$ will play the role of number
of coupled states $M l_1 = R$ in an effective thick wire
so that $l_c/l_1 \sim R$. Of course, this growth can continue
only up to $R < l_{12}$ where $l_{12}$ is
one-particle localization length in two dimensions and
$\ln l_{12} \sim {l_1} \gg 1$. While our results definitely show
the increase of enhancement with $R$ the power of growth is around
0.25 and is significantly less than 1. We attribute this difference
to the fact that the ratio $R/l_1$ was not big enough
(in  Fig.9 $ R/l_1 < 2.1$) and the asymptotic regime was not yet reached.
The further increase of $R$ is quite difficult since $l_c$
becomes comparable with the size of the basis.

In general our results confirm the relation (\ref{est}) but a more detailed
verification of this equation is still desirable.

Let us now discuss in more details the different consequences of
the result (\ref{est}). First we start from different dimensions $d$.
For $d=2$ the length $l_1$ in (\ref{est}) should be understood as
one-particle localization length in 2 dimensions. The
number of transverse channels $M$ is approximately equal to
$l_1$ so that finally $l_c \sim {l_1}^3$. For dimension $d=3$
an interesting situation appears below Anderson transition
for one particle \cite{D94}.
Indeed, it is possible to realize a random potential in which
{\it all} one-particle eigenstates are localized for the
hopping strength $V < V_c$
(a shift of mobility edge by interaction is not a very
interesting case). As a typical example let us consider
the Lloyd model with diagonal disorder $E_{n_1,n_2,n_3} =
\tan \phi_{n_1,n_2,n_3}$ and hopping $V$ on a cubic lattice,
where $\phi_{n_1,n_2,n_3}$ are random phases homogeneously distributed in the
interval $[0,\pi]$. In this case $V_c \approx 0.2$ and below this value
{\it all} states are localized. For two interacting particles in such random
potential the effective strength of hopping for a pair will be
strongly enhanced $V_{eff} \sim \sqrt{N} U$. Here $U$ is on site
(or nearby site) interaction and $N \sim {l_1}^3$ is the
effective number of states coupled by interaction. Since $l_1$
can be quite large near (but below) one-particle transition point
$V_c$ then two particles, even if characterized by repulsive
interaction, can be delocalized
when {\it all} one-particle states are exponentially localized \cite{D94}.
Another way to see this effect is to say that the pair feels the
disorder averaged over the size of the pair $l_1$
which gives a
strong effective decrease of disorder. Since in 3d delocalization
takes place for $V_{eff} > V_c$ generally there is no requirement
to have $l_1 \gg 1$ and it is not necessary to
take $V$ very close to $V_c$.
The condition
$V_{eff} > V_c$ gives the boundary of pair delocalization
$U {l_1}^{3/2}/V > 1$.

The appearance of delocalization for a pair in 3d leads to
quite interesting properties of energy spectrum.
Indeed, for particles located on a distance $r \gg l_1$ from
each other the effective interaction is exponentially
small ( $\sim \exp(-2r/l_1)$) due to the small overlapping of
one-particle states. Therefore, such states remain localized
while the delocalization will take place only for the states
with interparticle distance $r < l_1$. Since the localized states
with $r \gg l_1$ form an everywhere dense spectrum
this would mean that the continuous spectrum, corresponding to
a delocalized pair, is {\it embedded} into the pure point spectrum
of almost noninteracting one-particle states.

Generally speaking such kind of spectrum is unstable with
respect to small
coupling between quasi-degenerate levels. In the present case the coupling
is exponentially small but nevertheless it can change in principle
the structure of the spectrum. The physical reason of such
possible change can be understood in the following way.
The delocalized pair propagates in a random potential which acts as some
effective noise. This can increase the size of the pair
even if the matrix elements for transitions with $r=n_{-} \gg l_1$
are exponentially small.
Due to this noise the size of the pair will grow in time.
The rate of growth can be estimated as
$D_{-} ={n_-}^2/t \sim {l_1}^2 \exp(- 2 {n_{-}}/l_1)$.
This gives a logarithmically slow growth of the pair size
$n_{-} \sim (l_1/2) \ln t$. At the moment it is not quite clear what will
be the effect of the pair size growth on pair propagation in $n_+$.
At minimum, the displacement of the pair should become
slower than diffusive ${n_{+}}^2 \approx (n_1-n_2)^2 \sim t/\ln t $.
However, it is
quite possible that sticking at $n_{-} \gg l_1$ will produce a
more significant effect on the growth of $n_{+}$ since in the
region $n_{-} \gg l_1$ the matrix elements for transitions in
$n_{+}$ are also exponentially small. It is interesting to note that
even in the case of strong attraction between particles
the coupled state should be destroyed during the propagation in a random
potential. Indeed, during the displacement of the pair
disorder leads to transitions from the coupled state to
continuum leading to the destruction of pair. Usually, the destruction rate
is proportional to the squared amplitude of disorder
and this can make the life time of coupled state relatively short.
Contrary to this case the effective life time of a pair of repulsive
particles discussed above can be much larger since
$n_{-}$ grows only logarithmically with time.
In some sense the interference creates exponentially high barriers
which effectively push particles to stay together.
In quasi-one-dimensional case with $l_1 \gg 1$
the effects of slow pair size growth can also lead to
the appearance of logarithmic corrections in the expression of the
enhancement factor in (\ref{est}). For example,
we expect $l_c/l_1 \sim
l_1/(\ln l_1)^{\nu}$ with $\nu \sim 1$ for $M \sim 1$.
The effects of TIP
in 3d systems below Anderson transition when all one-particle eigenstates are
localized are quite interesting and at present, we try to
study them in numerical simulations with
effective 3d models \cite{BS}. Recently, an interesting
approach to the TIP problem in $d$-dimensions was introduced in
\cite{Imry}.

Up to now we discussed the effects of interaction only
for two particles. However, for solid state systems
the natural question is what will happens for a finite particles
density $\rho_{e}$. As it was discussed in \cite{TIP} the above
picture of TIP can be quite useful in the regime of
small density $l_1 \ll 1/\rho_{e} \ll l_c$. In this
case the interaction is mainly
reduced to interaction between two isolated particles.
If all the particles are separated from each other by a distance
$L \sim 1/\rho_{e} \gg l_1$ then the interaction is exponentially small,
all particles are localized and the current through such sample
is exponentially small. However, it is possible to have
another type of configuration when the particles are distributed by pairs
of size $l_1$. In this case pairs can easily propagate
on a distance $l_c \gg L \sim 1/\rho_{e}  \gg l_1$. Collisions of pairs will
go in a random way and will destroy interference effects for a pair.
These collisions will lead to delocalization and appearance of
finite conductivity in an infinite system.
It is interesting to note that it is enough to have
only one pair when all other particles are well separated
by the distance $L \gg l_1$. Then the collisions will allow to
transfer the charge through the whole sample.
However the above consideration,
based on (\ref{est}) and
being correct for particle energy at the center of the
band ($E \sim V$),  should be applied more accurately
for low energies near the ground state.
Indeed, as it was  discussed in \cite{TIP}
at low energies one should consider a transition
from a lattice to a continuous system in which
the enhancement factor should be proportional to
$l_c/l_1 \sim (k_{F} l_1) M$ since $k_{F} l_1$ determines the number of
independent components in a localized state (for $M \gg 1$
the factor $M$
should be replaced by $ k_F a_t$ where $a_t$
is the transverse width of the sample).
Near the ground state $\rho \sim k_{F}$ and it seems that
condition $l_1 \ll 1/\rho_{e} \sim 1/k_{F}$ implies that the
enhancement does not work at low energies. Due to that
at small densities there is no formal contradiction with the
results \cite{Tieri} according to which
repulsive interaction reduces the localization
length near the ground state.  To have a better understanding
of the situation at small $\rho_{e}$ a more exact analysis should
be carried out to obtain a more precise expression for $l_c$
in the continuous limit. In principle, the
average difference of energies for two repulsing particles
($k_F \sim 1/a=1$) on a distance
$r \gg l_1$ $(E_{\infty})$ and $r < l_1$ $(E_{l_1})$
is of the order of $\vert E_{\infty}- E_{l_1} \vert \sim U/l_1$
and is not very large for
large $l_1$. In fact this difference is
even less than the amplitude of disorder $W$
(we take the case of 1d Anderson model discussed in
\cite{TIP} with diagonal disorder in the interval $\pm W$
where near the center of the band $l_1 \approx 25 (V/W)^2$).
It is possible that for investigation of continuous limit
$k_F a \ll 1$ at low energy the approach used in \cite{Dor}
for two particles with strong attraction can be useful
after some extension.

The most interesting case with density $\rho_{e} \sim1$
formally cannot be analysed on the basis of
the result (\ref{est}) for TIP. However, it is possible to think that
interaction between quasi-particles can be studied
in the same way as for TIP and
that at small density of quasi-particles $\rho_q$
with $k_{F} \sim 1/a = 1$ ($l_1 \ll 1/\rho_q \ll l_c \sim {l_1}^2$)
conductivity will be not exponentially small for one-dimensional samples with
a size $l_{sam} \gg l_c \gg l_1$. In 3d for a "gas" of quasi-particles
the possible slow growth of the pair size should be
less important since collisions between pairs
give rise to destruction of interference and finite
conductivity
in the regime where all quasi-particles are localized.
Due to existence of exact connection between localization
in 1d and 1d disordered spin systems \cite{DFisher} it would be interesting
to understand possible manifestations of the analog of two
particles interaction for spin systems.

Finally, let us briefly discuss the possibilities of
application of the observed enhancement for explication
of large persistent currents observed in the experiments
with small metallic rings \cite{ring}.
Formally
the coherent localization length (\ref{est})
is strongly enhanced in presence of interaction.
Nevertheless, the direct estimates for the model of interacting
kicked rotators and numerical results (see Fig.1 and \cite{TIP})
clearly show that
the diffusion rate on the time scale $l_1 \ll t \ll l_c$
is not larger than the classical rate at
$t \ll l_1$. It follows that the time to cross
a sample will be not decreased by interaction.
However, the magnitude of persistent current depends not only
on the diffusion rate but also on the density of levels
which in principle can become very large for multi-particles
systems. Therefore, the possibility of enhancement of
persistent current due to interaction is still open and
should be studied in more details.

\section{Acknowledgments}
\label{sec:level5}

One us (F.B.) would like to thank the Lab. de Phys. Quantique,
URA 505 CNRS, Universit\'e Paul Sabatier for the kind hospitality
during his visit when this work had been started.
Computer assistance from Dr. Alberto Cominelli is   greatly
acknowledged.
Support from the Centre du Calcul Vectoriel
pour la Recherche (CCVR) Palaiseau, France is acknowledged.
It is our pleasure to thank Oleg Sushkov for valuable discussions.

\begin{figure}
\caption{
Dependence of second moments on time
in model $(1.1)$ with $k=7$, $K=kT=5$, $U=2$;
upper curve is $\sigma_+$, lower is $\sigma_-$. At $t=0$
both particles are at $n_1=n_2=0$, basis is
$-800 \leq n \leq 800$. For $U=0$ $\sigma_+(t) \approx 600$ for
large $t$.
}
\end{figure}

\begin{figure}
\caption{
Probability distribution
for two particles in the case of Fig.1 at $t=8 \times 10^4$.
Different contours show different probability levels.
}
\end{figure}

\begin{figure}
\caption{Probability distribution as a function
of $n_\pm = 2^{-1/2} (n_1 \pm n_2) $
for the case of Fig.2:
$P_+ (n_+ )$ (full line); $P_- (n_- )$ (dashed); dotted line is the
theoretical distribution $P_+^0 (n_+ )$ for $U=0$.}
\end{figure}

\begin{figure}
\caption{
Dependence of localization length $l^+$
(full circles) and $l^-$
(open circles) on one-particle localization length $l_1 = k^2/2$ for
$K=5$, $U=2$ and $4 \leq k \leq 8$. Full line shows dependence
$l_c = l^+ \propto l_1^2$, dashed line marks $l^- \propto l_1$.
}
\end{figure}

\begin{figure}
\caption{
Dependence of enhancement for $\sigma_{\pm}^{\infty}$
on $U$ for the model (2.1) with random rotation phases
(see section II), $k=4$. Error bars are obtained from $\sigma_{\pm}^{\infty}$
for 10 different realizations of disorder.
}
\end{figure}

\begin{figure}
\caption{
Dependence of $\sigma_{+}$ on time
for the model with finite interaction radius;
$k=5.7$, $K=5$, $U=\pi$; $R=1$ (full curve), $9$ (dotted), $33$ (dashed).
Initial conditions are as in Fig.1, basis is
$-500 \leq n \leq 500$. The noninteracting case $U=0$ has $\sigma_{+} \approx
250 \approx {l_1}^2 $ (see Fig.3 in [10]).
}
\end{figure}

\begin{figure}
\caption{
Same as in Fig.6 but for $\sigma_-$.
}
\end{figure}

\begin{figure}
\caption{
Probability distribution as a function of
$n_\pm = 2^{-1/2} (n_1 \pm n_2 )$ for the case
of Fig.6 and $R=9$: $P_+(n_+)$ (full line),
$P_-(n_-)$ (dashed line). $P_+^0 (n_+) $ (dotted line)
is the theoretical distribution for $U=0$.
}
\end{figure}

\begin{figure}
\caption{
Dependence of $\sigma_{+}^\infty$ (full circles)
and $\sigma_{-}^\infty$ (open circles) on interaction radius $R$;
$k=5.7$, $K=5$, $U=\pi$.
}
\end{figure}

\end{document}